\RequirePackage{fix-cm}
\documentclass[twocolumn]{svjour3}       
\smartqed  
\usepackage{graphicx}
\usepackage{hyperref}
 \journalname{Experimental Astronomy}
\usepackage{comment}
\newcommand{\crires}{\ensuremath{\mathrm{CRIRES}^\texttt{+}}}

\begin{document}

\title{End to end simulators: A flexible and scalable Cloud-Based architecture.}

\subtitle{Application to High Resolution Spectrographs ESPRESSO and ELT-HIRES.}

\author{M. Genoni  \and M. Landoni \and G. Pariani \and M. Riva \and A. Bianco \and G. Li Causi \and T. Marquart \and F. A. Pepe \and A. Marconi \and E. Oliva}

\institute{M. Genoni (corresponding Author)\at
              INAF-Osservatorio Astronomico di Brera \\
              Tel.: +39-02-72320477\\
              \email{matteo.genoni@inaf.it}
}

\date{Received: 27 March 2020 / Accepted: 13 August 2020}

\maketitle

\begin{abstract}
Simulations of frames from existing and upcoming high-resolution spectrographs, targeted for high accuracy radial velocity measurements, are computationally demanding (both in time and space).
We present in this paper an innovative approach based on both parallelization and distribution of the workload. By using NVIDIA CUDA custom-made kernels and state-of-the-art cloud-computing architectures in a Platform as a Service (PaaS) approach, we implemented a modular and scalable end-to-end simulator that is able to render synthetic frames with an accuracy of the order of few cm/sec, while keeping the computational time low. 
We applied our approach to two spectrographs. For VLT-ESPRESSO we give a sound comparison between the actual data and the simulations showing the obtained spectral formats and the recovered instrumental profile.
We also simulate data for the upcoming HIRES at the ELT and investigate the overall performance in terms of computational time and scalability against the size of the problem. In addition we demonstrate the interface with data-reduction systems and we preliminary show that the data can be reduced successfully by existing methods.


\end{abstract}

\section{Introduction}
\label{intro}
End-to-End instrument models (E2E) are numerical simulators which aim to simulate the expected astronomical observations starting from the radiation of the scientific sources to the raw-frame data produced by instruments. Synthetic raw-frames can be ingested by the Data  Reduction Software (DRS) to be analyzed in order to assess if top-level scientific requirements, such as spectral resolution, SNR, Radial Velocity (RV) accuracy and precision, are satisfied within the specific instrument design.

More and more heterogeneous and complex instrumentation, especially for future extremely large telescopes, are stressing the need of such tools.  In fact, E2Es have been valuable software exploited in many types of astronomical instruments for different purposes. Specifically, they have been used in design phases to optimize and improve specific hardware components and parameters \cite{Ref-LiCausi-1}, or for early verification of instrument performance (see \cite{Ref-Ilocater2}). In addition, they helped the verification of performance during instrument integration by comparing theoretical and real data such as image quality, spectral resolution and spectral format (\cite{Ref-Jarno}). From the scientific point of view, they have been extensively exploited for assessing the feasibility of particularly challenging observations (see e.g. \cite{Ref-WEBSIM}). Furthermore, instrument simulators are systematically exploited to aid both Data Reduction (\cite{Ref-Xshooter}) and Data Analysis Software development, as well as for testing and verifying existing data reduction pipelines. From the operation point of view, they are also capable to drive calibration procedures and observation plans. Considering modern and future extremely precise radial velocity spectrographs, RV error budgets (see \cite{Ref-Ilocater1} and \cite{Ref-Halverson}) will benefit of accurate simulations, when identifying and characterizing the different noise sources (like, for example, telescope windshake and fiber scrambling) and instrument systematics (e.g. bench and dispersers thermo-mechanical stability) expected to set the limits in Doppler measurements.
In most of the cases, simulators are targeted and designed for a single instrument and so tightly tailored for a specific application although, recently, efforts have been put in the direction of developing general purposes simulator (see e.g. \cite{Ref-Stuermer}).

Moreover, a key aspect in the context of high precision RV spectrographs (e.g. HARPS and ESPRESSO, see \cite{Ref-Udry} and \cite{Ref-Pepe-ESPR} respectively) is the fidelity of the simulated frames, required to avoid spurious numerical errors.
In particular, if one wants to simulate a synthetic echellogram of such instrumentation, it has to guarantee that the computation of the flux (on each pixel and for each wavelength) has an accuracy such that the residual shift on the photocenter of each resolution elements, in RV units of cm/s, must be lower than both the global instrumental precision and errors from all noise sources (and instrument systematics), which are typically of the order of 2-10 cm/s.
In term of pixel scale, the photocenter residual shift shall be lower than $1/10^{5}$ of pixel, thus turning directly into a challenging computational problem. For instance, for a R=100000 multi-fiber spectrograph, like ESPRESSO or ELT-HIRES, the numerical simulation of high signal-to-noise (SNR) echellogram would easily exceed a total computation time of about 100 days with a single core machine. 
For this reason, it is necessary to adopt both parallel and distributed computational architecture in order to guarantee high precision and reasonable required time to complete a simulation. 

We present here a modular E2E simulator designed adopting an innovative architecture strongly based on both parallel computing with CUDA and Cloud Computing services (see e.g. \cite{Ref-landoni19a} and \cite{Ref-landoni19b}). 
We exploited our tool to simulate two different high RV precision spectrographs, both single fiber-fed, ESPRESSO, and multi fibers ELT-HIRES, aiming to demonstrate the applicability of our solution and its flexibility in terms of underlying different instrumentation.
The final goal of this simulator is its full integration with a DRS pipeline, in order to close the loop (from the 1D input spectrum to 1D extracted one). 

ESPRESSO is the Echelle SPectrograph for Rocky Exoplanets Stable Spectroscopic Observations for the Very Large Telescope (VLT) observatory. It covers the full visible spectrum, its resolving power (R) can reach 240000 and it has an instrumental RV precision of 10 cm/s, for further details see \cite{Ref-Pepe-ESPR}.
ELT-HIRES is the planned High Resolution Spectrograph (\cite{Ref-Marconi}) for the future Extremely Large Telescope (ELT). In its current design, it foresees 8 spectrometers working in the 0.4-1.8 um bands, with R=100000 and instrumental RV precision of 1 m/s. One of its main scientific goal will be exoplanet atmospheric characterization.
For the former, we will compare the simulated image and the actual one recorded by the instrument while for the latter we present the expected echellogram in the Z band in spectro-polarimetric mode.

This paper is organized as follows. In the Section 2 we present the E2E simulator architecture by focusing on the main relevant modules (image simulation module and spectrograph unit) while in Section 3 we show the computation architecture adopted to fulfill the goals here described. Finally, we discuss the application of the simulator to ELT-HIRES and ESPRESSO in Section 4.

\section{Design Philosophy and Architecture}
The philosophy on which the whole synthetic image (spectral format) simulation is based, reflects the physics behind it; i.e. it is built as the sum of the single fibers image (or the entrance slit image, for slit-fed spectrographs) at each wavelength projected onto the detector. The wavelength grid step, used in the simulation, can be chosen accordingly for the specific required accuracy.
Thus, being the simulation deeply parallelizable, the development and implementation of a general purpose and scalable E2E, is accomplished by setting the proper computational architecture. 
Indeed, the adoption of parallel and cloud computing paradigm (see sec. \ref{subsec:Image_Sim_Module} and sec. \ref{sec:CloudComputing}), is fundamental for enhancing the scalability of the simulator, allowing to tune the computational power/resources according to the different kind of simulations that can be performed in relation to different and possibly growing complex instruments.

\begin{figure*}
\centering
{\includegraphics[width=0.85\textwidth, height=0.85\textheight, keepaspectratio] {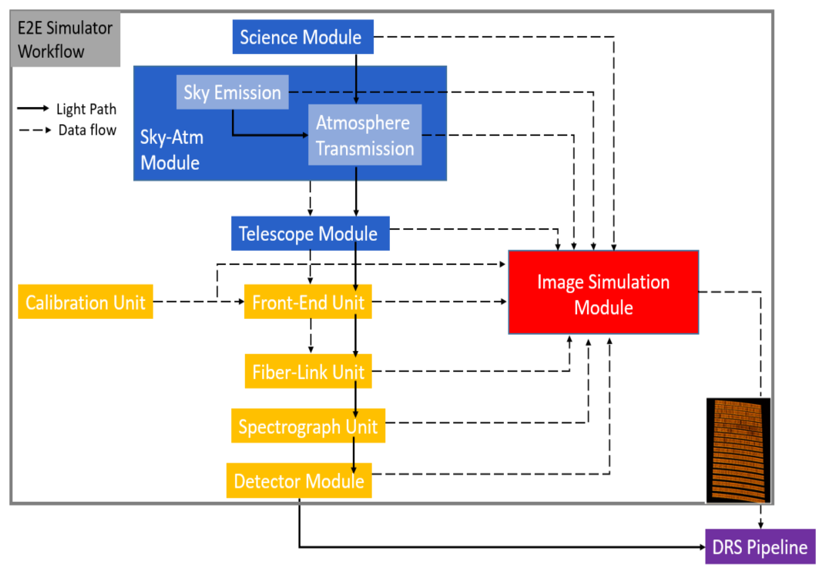}}
\caption{E2E simulator workflow schematic description, for fiber-fed spectrographs like ESPRESSO and ELT-HIRES.
The solid arrows represent the real light path, while the dashed arrows show: the simulation data-flow, how the different modules and units are interfaced and their connection to the simulator core, which is the Image Simulation Module. The synthetic frame produced by the simulator is ingested by the DRS pipeline.
Yellow blocks are units of the Instrument Module, while blue blocks are related to simulation modules independent from the specific instrument.}
\label{fig:E2E_1_Glob_Arch_03}
\end{figure*}

Likewise, the End-to-End simulator architecture is highly modular, composed by different modules (each one with specific tasks) units and interfaces, as described in the schematic workflow of Fig. \ref{fig:E2E_1_Glob_Arch_03}. 
Modularity and flexibility are key points, in the definition of interfaces among different modules or units, to allow our system for being scalable and adaptable to different kind of spectroscopic instrumentation. Modules and Units are characterized by the main tasks for which they are in charge of, required inputs and expected outputs (in a specified format).

For example, the Science Object Module has the purpose to generate a synthetic spectrum, at suitable resolution, of a specific science source, obtained by a set of parameters (e.g. spectral type, magnitude, red-shift and radial velocity for stars) or by loading the spectrum from a user-defined library. The output is a FITS table containing the flux, in units of photon flux, at each wavelength. 
The Front-End Unit (which belongs to the Spectrograph Module) is in charge to predict the light distribution at the Fiber-Link entrance focal plane and to evaluate its throughput. In the case of ELT-HIRES, this focal plane harbors the dicing-unit, which samples the field with a micro-lens array and separates the portions of the field that feed the corresponding fibers in the fiber bundle (see \cite{Ref-Oliva} for further optical design details). 
The Front-End Unit and Fiber-Link Unit are examples of units which can be interchanged, according to the specific spectrograph layout, i.e. single object fiber-fed or multiple objects fiber-fed (i.e. mos), while in the case of slit-fed spectrograph the Fiber-Link Unit can be substituted by a Slit-Unit which models the slit geometry and determines the slit-loss. This will be shown in a future paper focused on the characterization of our simulator for the upcoming slit-fed spectrograph ESO-SOXS (see e.g. \cite{Ref-Schipani}).

We focus hereafter on the two most relevant parts of the simulator for which an innovative approach for their implementation is necessary to reach our aims: the Spectrograph Unit, designed for exploring different levels of physical approximations, and the Image Simulation module, draw to deeply exploit high performance computing paradigms.

\subsection{\textit{The Spectrograph Unit}}
This Unit has the purpose to simulate the physical effects of the different optical components of the spectrometer for predicting the spectral format at instrument focal plane, the throughput and the light distribution of the camera entrance pupil (for estimating second order effects on the object spectrum). 

\begin{figure*}
\centering
{\includegraphics[width=1\textwidth, height=1\textheight, keepaspectratio] {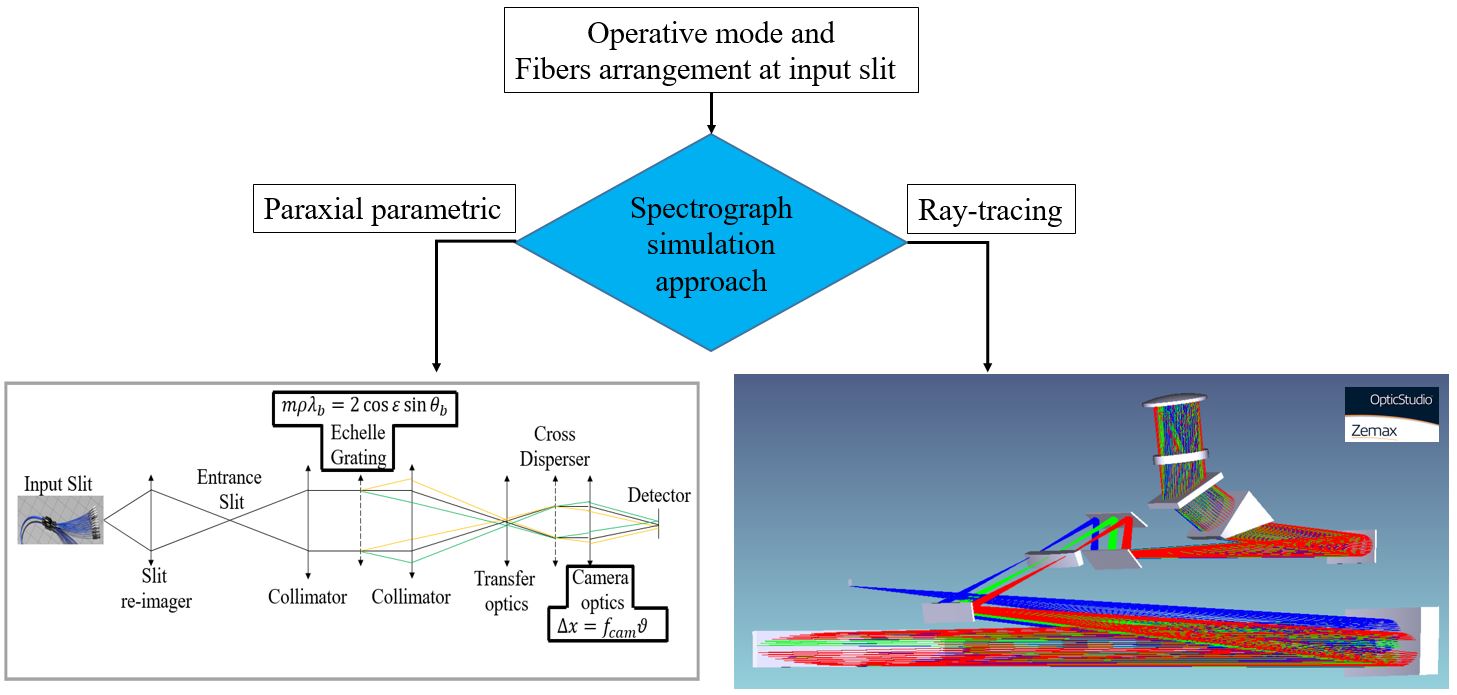}}
\caption{\footnotesize
Schematic description of how the simulator can handle the two approaches, for modeling the spectrograph optics. The Parametric mode, left path, build a paraxial model relying on the equations which physically characterize the optics. The Ray-Tracing mode, right path, exploits ray-tracing optical files, usually built in commercial software packages like Zemax-OpticStudio.}
\label{fig:E2E_16a_Parax_vs_RayT_2}
\end{figure*}

One of the peculiar and innovative feature of the simulator working mode we developed is that the spectrograph can be modeled using two different alternative approaches: the Parametric Version and Ray-Tracing Version. The former is based on a physical parametric model (see for example \cite{Ref-GenoniPPM}), built with the physical equations and relations which characterize the optical elements, while for the latter we developed an ad-hoc software wrapper to interface with commercial optical design CAD (e.g. Zemax/OpticStudio). A schematic view that illustrates how the simulator can handle the two approaches is shown in Fig. \ref{fig:E2E_16a_Parax_vs_RayT_2}.

In the Ray-Tracing mode, the Unit takes as input the optical design files (built with the commercial software packages like Zemax-OpticStudio) and uses tailored API-based software developed to extract the required outputs (spectral format at instrument focal plane, the throughput and the light distribution of the camera entrance pupil). 
Instead, the Parametric Version works without the need of the optical design files and requires only geometrical parameters (such as dispersing elements working angle, optical elements working F-Number) to recover the expected spectral format required for rendering the image.
The Parametric Version allows to execute a complete run of the simulator, even if the optical design of the instrument is only at a preliminary phase and the detailed optical ray-tracing files are not available. 
This approach is very useful for quick parametric evaluations and analyses concerning different possible spectrograph design choices and architectures. Furthermore, it allows to draw considerations and comparison regarding the orders curvature, lines tilt, fibers alignment which are relevant aspects for the data reduction software. 

In addition, this unit is also in charge to evaluate the optical Point Spread Function (PSF) obtained at the level of the focal plane for each wavelength. In particular, the instrument PSF modeling is described within two levels of approximation. The former is related to aberration and diffraction coming from the optics of the spectrograph and it can be carried out analytically (adopting an asymmetric Gaussian function, to take into account different first order aberrations) or by using the interface with ray tracing software to recover the complete PSF. The latter takes into account the diffraction spikes coming from the obstruction present, for example, in Schmidt cameras proposed in the current ELT-HIRES spectrograph optical design (see \cite{Ref-Oliva} and \cite{Ref-LiCausi-2} for a deep analytical description and implementation of spikes effects).

The output of the Spectrograph unit is a database composed by spectral format data, PSF maps and throughput for each projected spectral fiber image (SFI) on the detector. Spectral format data comprise the diffraction order number, wavelength and spanned spectral range, fiber identification number, centroid coordinates, and size (i.e. geometrical full width in main and cross dispersion directions) of each SFI image. The grid step of SFI is set to be a fraction of the physical spectral resolution to avoid aliasing effects in the rendered image (for details see sub-Section 2.2). 

The database, which could be either a plain text file or a JSON no-SQL database for higher scalability (e.g. MongoDB, see \cite{Ref-kanade}) is thus queried by the Image Simulation Module (see next section). 
This approach allows to decouple the part of the simulation related to the instrument and the one related to the numerical computation, allowing the image simulation kernel to be adapted to different instruments. 
Moreover, for highly stable spectrographs, like ELT-HIRES and ESPRESSO, the database can be computed only once and used for different simulation scenarios, since the spectral format and PSF maps are not expected to evolve and change.
In this case, the Image Simulator provides an illumination profile for each fiber that, since normalised, can be multiplied only at the end simulation chain by the number of photons received in the wavelength range spanned by the SFI.

\subsection{\textit{Image Simulator Module}}
\label{subsec:Image_Sim_Module}
This module is the core of the whole system. This piece of software is responsible for rendering the distribution of the photons of each single spectral fiber image as should be detected at the level of the detector. The final image of the echellogram is obtained as a superposition of individual spectral images of the fiber sampled on a proper wavelength grid. Those are estimated as a surface integral of the convolution between the spatial profile of the fiber and the PSF, which is considered as the optical aberrations plus its diffraction component, computed on a sub-pixel sampling.
Then photon noise and all other specific detector noises (e.g. dark current, RON, pixels non-uniform response and cross-talk) are added on the final synthetic echellogram. 

\begin{figure*}
\centering
{\includegraphics[width=1\textwidth, height=1\textheight, keepaspectratio] {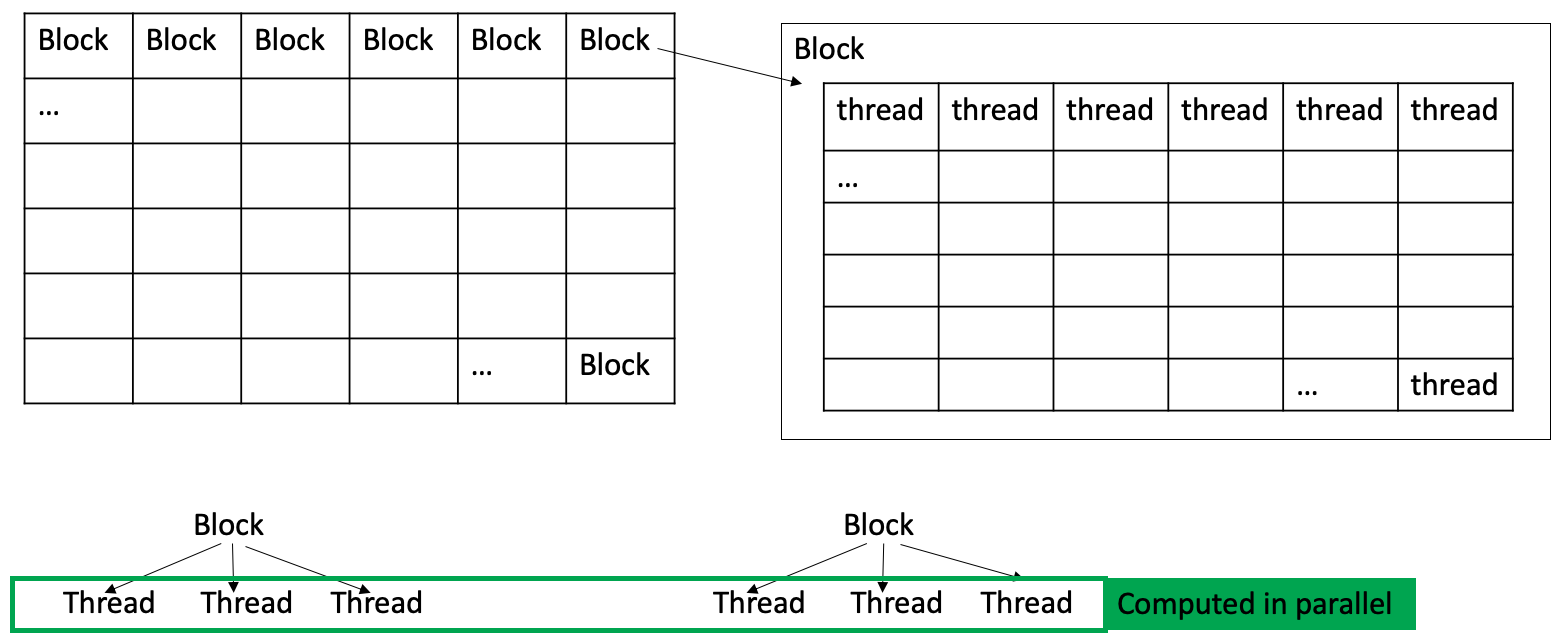}}
\caption{CUDA-based approach for the computation of the single fiber image. Each Block is in charge to evaluate the flux within each pixel by using 100 threads for the computation of the convolution between the geometrical image and the PSF.}
\label{fig:E2E_CUDA}
\end{figure*}

For a high resolution and high RV precision spectrograph, the photocenter of each single image on the detector shall be reconstructed with high accuracy, to avoid residuals, in terms of radial velocity, higher than the errors due to the individual noise sources and instrument systematics. 
This accuracy is necessary for a simulator that shall be reliable for a subsequent introduction in the simulation of noises and systematics for characterizing their impact on the final instrumental and scientific performance.
In particular, for a R=100000 spectrograph with a spectral sampling of 3 pixels, the computation of the flux (on each pixel for each wavelength) shall have an accuracy such that the induced shift on the resolution element photocenter must be lower than $0.1 nm$ (about $1/10^{5}$ of a pixel), meaning an error of 1 cm/s for the single spectral line.
This requirement translates directly into an intensive computational problem. For example, in the case of ESPRESSO, the image rendering of a single fiber requires on a single core machine about $\sim$ 70 sec (Intel Xeon X5650 @ 2.67 GHz) giving a total of $\sim$ 80 days to compute a full simulation of the echellogram. Clearly, it is necessary to find a way to reduce the time in order to make the simulator a valuable tool among the various design phases. 

For this reason, we have developed a heavily parallel computation procedure that uses an approach based on the parallel computing paradigm CUDA by NVIDIA for the evaluation of the flux at each pixel for each resolution element.

In details, the computation of each fiber image is performed on a grid of N $\times$ N pixels as depicted in the schematic of Fig. \ref{fig:E2E_CUDA}. 
In particular, for each sub-pixel, an independent CUDA thread evaluates the flux by numerically integrating the photon distribution derived from the convolution of fiber profile and the PSF. Threads that belong to the same Block (a CUDA abstraction that represent a group of thread that, in our case, is assigned to each pixel) sum their result at the end of their computation on a shared memory region on the GPU. In our architecture, this block memory region models the flux received within each considered pixel. This approach allows to run in parallel 10.000  thread on common NVidia GPU. 

The grids of wavelengths and pixels used for simulations are free parameters. It is known that an approach like the one presented in this work suffers from aliasing on the final image due to sampling, both in wavelength and pixel scale. We explored the parameter space and we found that using a grid of 10x10 pixels sampled at 0.1 pixels with a wavelength step of $\sim \frac{1}{10}$ of the actual spectral resolution is enough to reach the requirement on reconstruction accuracy, while maintaining aliasing effect just below the photon noise level. This result is shown in Figure 4, where we compare the photon noise limit on a simulated ESPRESSO flat-field frame with reasonable flux level ($\sim$ 40.000 ADU px$^{-1}$).

Furthermore, it is important to note that an alternative approach to the method presented in this paper would involve the direct ray-tracing of individual photons drawn randomly from the input spectrum, assuming an illumination distribution of the entrance fiber, to non-integer pixel positions and then binning to physical detector pixels. This kind of simulation has to advantage to be fast, able to rendering high-SNR images in few hours, but it assumes an analytical distribution on the input image and PSF and requires to simulate distinct images for each underlying spectral distribution (science object, flat field, etc.). 

A trade-off between the two methods can be performed considering the number of photons in the image and the number of SFI required to complete the simulation. As a rule of thumb, when considering  ESPRESSO\footnote{see the latest ESPRESSO Exposure Time Calculator at https://www.eso.org/observing/etc/bin/simu/espresso.} if one observes a target with $m_{v} \sim 16$ for 15 mins, the number of collected photons is of the order of $\sim 10^{7}$, comparable to the required number of spectral fiber images. However, the reached SNR is $\sim$ 10 and the accuracy on RV is $\sigma_{v} \sim$ few m s$^{-1}$. On the other hands, when considering high SNR spectra ($\sim$ 500) aiming to reach $\sigma_v \sim$ 10 cm $^{-1}$ ($m_v \sim 5$ and exposure time of few hundreds seconds) the number of photons easily exceed $10^{11}$ and the required amount of computational time for the two methods become similar. However, in the case of ray-tracing, the simulation must be repeated for each input spectral distribution (object and calibrations). For very high SNR (like the one required for precise RV measurements) our method appears to be more scalable than the one based on ray-tracing. Nevertheless, the consideration done about parallelism with the adoption of CUDA and Cloud Computing (see Section 3) can be adapted also for the ray-tracing approach by properly modifying the \textit{Image Simulator module} computational architecture.  
\\

We tested our method by developing an ad-hoc \texttt{C++} source code based on CUDA 11 SDK that has been executed using an NVidia Tesla K80. We obtained an overall median computation time of about 0.17 $s$ for each SFI, which guarantee a good vertical scalability (same machine), while maintaing the photocenter reconstruction error below a fraction nm as required. 
Moreover, the optimization of the computational cost shall be improved not only by using parallel execution with CUDA on the same machine, but also with the implementation of proper architecture to distribute the computation among many nodes within cluster (horizontal scalability). 
For example, in the case of ELT-HIRES one could distribute the simulations across different nodes, along the network, to compute in parallel the echellograms rendering of the 8 different spectrographs. Alternatively, with more granularity, one might use many nodes to distribute the computation of both spectrographs and orders within each instrument. This horizontal scalability requires to develop an architecture that auto-scales against the size of the problem with proper logic to distribute the workload and orchestrate the whole computation. A natural choice within this framework is the adoption of Cloud Computing (see e.g. \cite{Ref-landoni19a} and \cite{Ref-Williams}) since it gives both computational power and services tailored to horizontally scalable applications, that share  underlying philosophy like ours.

For our purpose, we exploited the capability offered by Amazon Web Services (AWS) since it combines both computational power availability (especially in terms of instances equipped with GPU) and reasonable cost. We detail in the next section the adopted architecture in our simulator.

\begin{figure*}
\centering
{\includegraphics[width=1\textwidth, height=1\textheight, keepaspectratio] {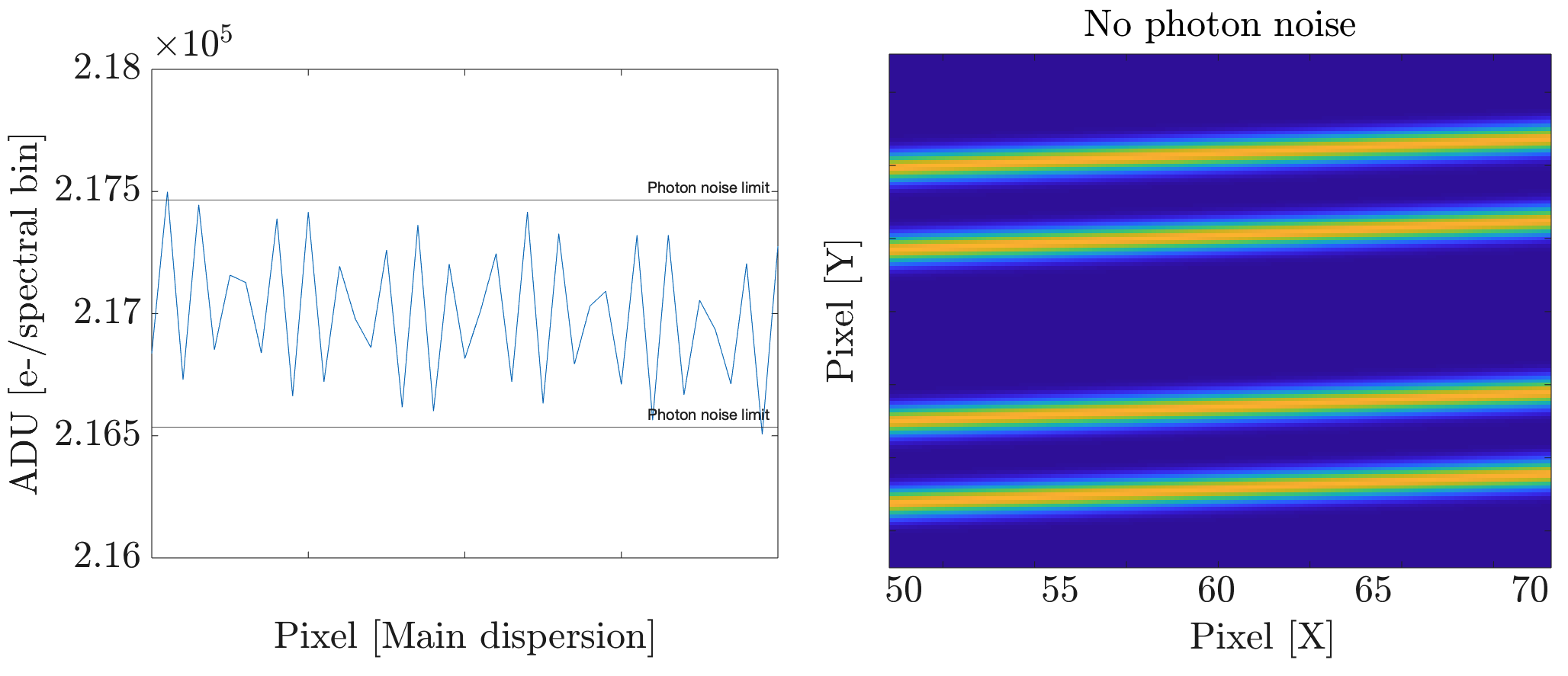}}
\caption{\footnotesize
Aliasing level, compared to the photon noise, for a flat field the blue arm of ESPRESSO with 10 images per resolution element (mean spectral resolution in the blue arm $\Delta\lambda = \frac{\lambda}{R} = 0.02$ \AA, mean wavelength grid of 0.002 \AA, mean flux value of $\sim$ 40000 e$^{-}$ px$^{-1}$). Left panel: extraction of a portion of flat field in ADU. The two vertical bars indicate the level of pure photon noise.  Right panel: 2D frame (zoom) of the same flat field before adding contribution from photon noise.}
\label{fig:E2E_CrossCheck}
\end{figure*}

\section{The Cloud Computational Architecture}
\label{sec:CloudComputing}
\begin{figure*}
\centering
{\includegraphics[width=0.95\textwidth, height=1\textheight, keepaspectratio] {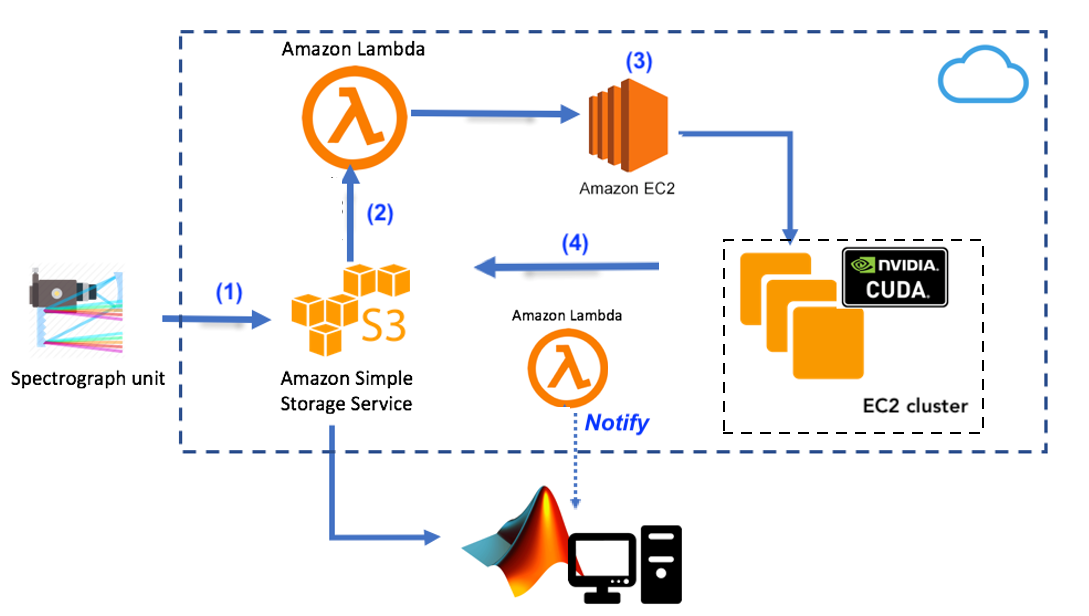}}
\caption{\footnotesize
Cloud architecture implemented for our E2E CUDA-based simulator. It foresees the use of Amazon \texttt{EC2} as Compute engine framework, Amazon \texttt{S3} for storing the data and the new \texttt{Amazon Lambda} service for automatically orchestrate the simulations.}
\label{fig:E2E_23_Cloud}
\end{figure*}

In Fig.\ref{fig:E2E_23_Cloud} we report the Cloud architecture based on AWS that has been implemented for the E2E simulator and subsequently exploited for the simulations discussed in this paper. 

The solution foresees the use of three main services offered by the platform: \texttt{Elastic Cloud Computing (EC2)} that basically offers virtual machines in the Cloud equipped with NVIDIA GPUs and accessible from the Internet, \texttt{Simple Storage Service (S3)} to store the data and \texttt{Amazon Lambda}, a computer service that runs used defined functions in response to some events triggered in the cloud, to orchestrate workload in the cluster. 

In details, the architecture follows this logic when a new simulation for a spectrograph is required. As soon as the Spectrograph units produces its database, a file containing the parameter of the simulation and the path of the database (if using a plaintext file) is uploaded to a specific bucket (aka, folder) in S3. 
When the file is upload (see Step (1), in Figure \ref{fig:E2E_23_Cloud}), a Lambda function is triggered (Step 2). This function checks the size of the problem by quickly analyzing the database and fires up an elastic cluster of EC2 instances, which number is defined accordingly to the dimension of the simulation. It also instruct each EC2 instance on which portion of the database (Key Partitions in the case of a non-SQL database) shall be analyzed to produce the relative fiber image.
These instances, when the computation of their portion of simulation is completed, upload their result as binary file on an S3 bucket. Finally, a second Lambda function notify the user when the last simulation chunk has been completed and, at this point, data can be retrieved locally for processing (e.g. with MATLAB).

\section{Results}
\label{sec:Results}
Here we present different simulation examples of our E2E tool, implemented for two high-resolution spectrographs ESPRESSO and ELT-HIRES. 

\subsection{\textit{ESPRESSO: Real data vs Simulations Comparison}}
\label{subsec:Results_ESPRESSO}
The Echelle SPectrograph for Rocky Exoplanets Stable Spectroscopic Observations (\cite{Ref-Pepe-ESPR}) is fiber-fed cross-dispersed echelle spectrograph placed in vacuum and in a thermally stabilized environment. 
It is installed at the Combined Coud\'{e} room of the ESO 8.2 m Very Large Telescope (VLT) observatory; it can be fed by any of the different Unit Telescopes (UT) of the VLT observatory through four optical Coude paths and its Front-End Unit.  
It can works in three different operative modes according to the used optical-fiber aperture: ultra high resolution (UHR, R=240000), high resolution (HR, R=120000), medium-resolution (MR, R=60000, this is that can also collect the light from the four telescope simultaneously). To enable these resolving powers, keeping the size of optics within manufacturing capability, the instrument implement pupil slicing before the entrance into the spectrograph. This means that each diffraction order has 4 spectral trace onto the detector, 2 for the target object fiber and 2 for the sky/calibration fiber.
We exploited real frames and data from ESPRESSO to test and to verify the reliability of our simulator outputs, run in the ray-tracing approach using an optical design file of the spectrograph to retrieve spectral resolution elements positions and PSF maps.

\begin{figure*}
\centering
\resizebox{\hsize}{!}
{\includegraphics[width=1\textwidth, height=0.7\textheight, keepaspectratio] {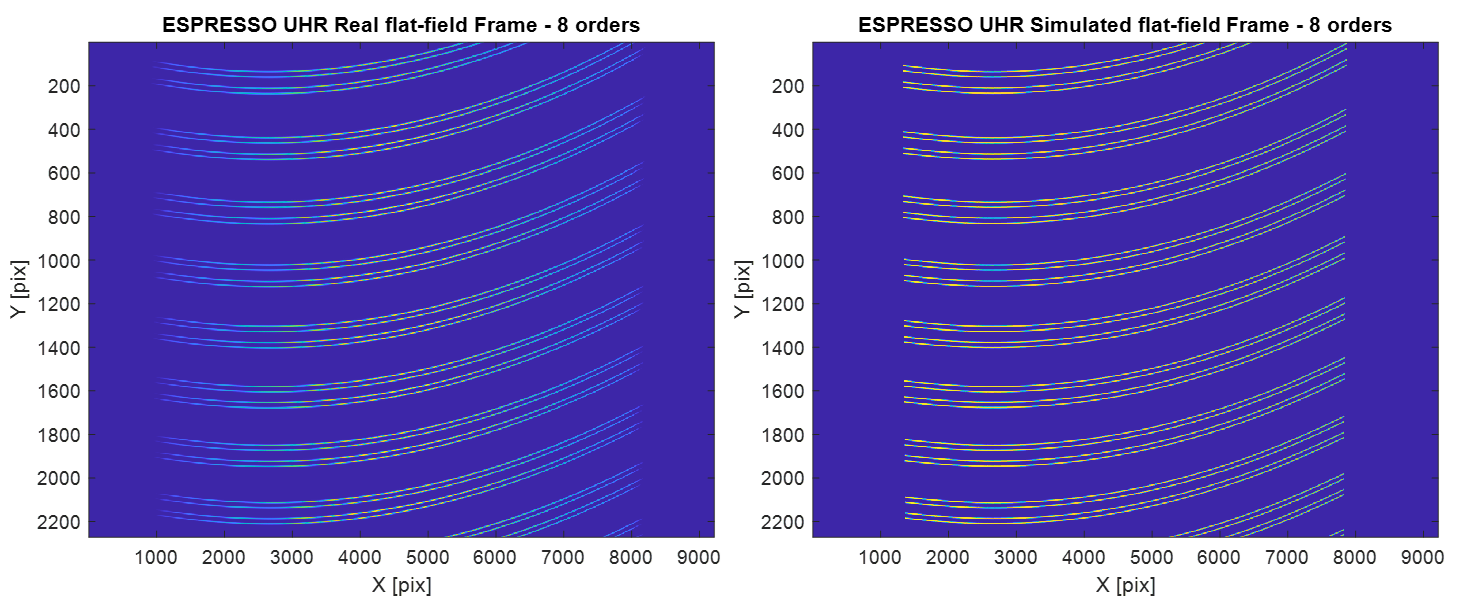}}
\caption{\footnotesize
Comparison of real vs simulated flat field frame of ESPRESSO, in UHR mode. 8 diffraction orders are shown to enhance visibility. 
\emph{Left}: Real frame; the intensity variation along each order due to the blaze function is clearly. The cut of orders along the main dispersion direction is due to a cut introduced by an optical component of the spectrograph.  
\emph{Right}: Simulated frame; the blaze function distribution and other optical throughput terms have not been emulated. The orders width is lower with respect to the real frame just because of the choice on the wavelength range spanned per each order.}
\label{fig:ESPRESSO_FlatField}
\end{figure*}

To test our E2E model, we produced a flat field in the UHR mode, without taking into account the blaze function distribution, and other optical throughput terms, in each diffraction order to enhance the contrast and visibility at orders edges. The comparison of the simulated versus real flat field, presented in Fig. \ref{fig:ESPRESSO_FlatField}, shows that the two spectral formats are geometrically consistent in terms of order curvature, order separations (along the vertical axis in the figure) and the 4 slices trace per each diffraction order. 
We extracted vertical cuts along different position along the main dispersion direction (X direction in the figure), through which we compared the simulated versus real orders and slices trace location and separation; the simulated values are in agreement with real ones within maximum deviation of about 1 pixel. This is not contradicting the requirement of computational accuracy explained in the Introduction and Sec. 2, since it is along the cross-dispersion direction, thus not affecting the radial velocity precision.

We also generated a Fabry-Perot calibration spectrum injected in both fibers of the UHR mode. 
The aim of this simulation is to check and compare the non-perfect alignment of the two half-slices image. The misalignment retrieved from both the simulated and the real frames is about 0.85 pixels, confirming the reliability of the simulation.
%
%

\begin{figure*}
\centering
{\includegraphics[width=0.8\textwidth, height=0.8\textheight, keepaspectratio] {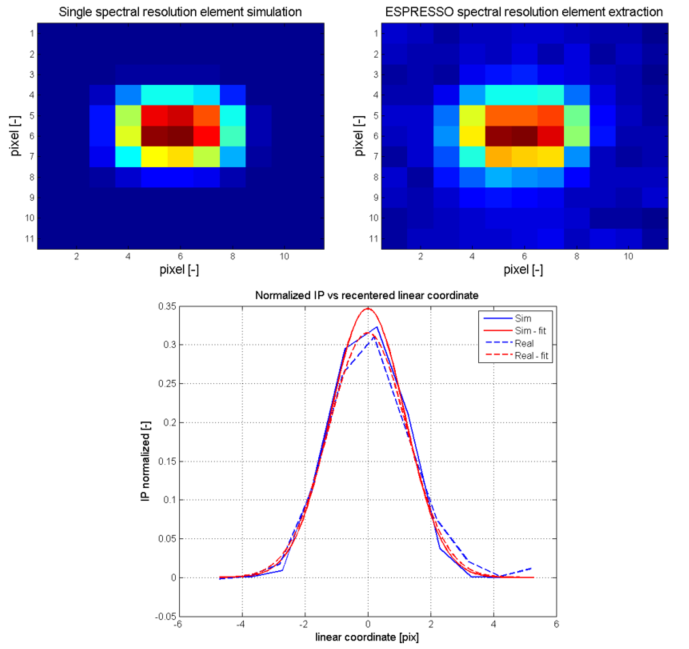}}
\caption{\footnotesize
Spectral resolution element light distribution and Instrumental profile comparison in UHR mode.
\emph{Top}: Qualitative comparison of a single Spectral line flux distribution. On the left a simulation, on the right spectral resolution element extracted from a real frame with ThAr lines. 
\emph{Bottom}: Normalized Instrumental profile comparison - the real and simulated profile (blue dashed and line)  show a good matching, the FWHM retrieved from Gaussian fits (red dashed and line) differ of about 0.004 pixels.}
\label{fig:ESPRESSO_IP}
\end{figure*}

Furthermore, we investigated the reconstruction accuracy of the Instrumental Profile (IP) simulated by the tool with the actual one recorded by the instrument. The IP is the image line profile of an unresolved spectral feature produced by the spectrograph.
This must be properly characterized in order to guarantee high accuracy wave-calibration and then achieve high RV performance.
We simulated different unresolved spectral resolution elements in different locations across the echellogram and we compared them with the ones from real Thorium-Argon (ThAr) frames.
For each of them, the flux distribution is first binned in spatial direction, and then fitted with a Gaussian profile to recover the full width half maximum (FWHM) values, as shown in Fig. \ref{fig:ESPRESSO_IP}. 
We restricted our analysis on the FWHM, and not on the overall lines shape, because the simulation does not yet include some features (like non-perfectly uniform illumination, thermal instabilities, etc.) which has an impact on the lines spreading. We are performing further comparisons and analyses, that will be presented in a future work. 
The mean FWHM deviation of the different simulated unresolved spectral lines with respect to the ones extracted ThAr is lower than 0.004 pixels, with a standard deviation of $\sim$ 0.001 pixels.

Finally, for what concern the computational performance, we were able to synthesize the echellograms of ESPRESSO here presented in just $\sim$ 5 hrs with a cluster of 4 \texttt{g3s.xlarge} EC2 instances (NVIDIA Tesla M60) with a total cost of roughly $\sim$ 5 USD. The use of these kind of instances, allows to reach a reasonable trade off between speed, computational power and overall cost while guaranteeing a strong availability in the unsold capacity stock on the AWS platform (Spot Instances, see \cite{Ref-landoni19a}).
In a forthcoming paper, we will show how the simulated frames from our model are ingested by the ESPRESSO DRS to explore different simulations scenarios.

\subsection{\textit{ELT-HIRES: Raw frames rendering and Data Reduction Software first tests}}
\label{subsec:E2E_7_Results_RAW_DRS}
The High Resolution Spectrograph for the Extremely Large Telescope, named ELT-HIRES (\cite{Ref-Marconi}), has a functional architecture which is highly modular. It foresees 8 independent fiber-fed cross-dispersed echelle spectrographs, each optimized for a specific wavelength band in the total spectral coverage from 400 nm to 1800 nm.
The instrument will be equipped with several fibers bundles (both for science target and sky), which are vertically re-arranged to create a pseudo slit at each spectrometer entrance. 
Different fiber bundles allow to achieve distinct resolution capabilities and observing modes. In the proposed instrument concept, these are: throughput maximization, spectral (RV) accuracy maximization, spatially resolved information (IFU), spectro-polarimetry and multi-objects medium-resolution spectroscopy.

ELT-HIRES is currently in pre-phase-B and in order to aid the Data Reduction Software design, we used our E2E simulator to explore the re-usability/applicability of existing DRS.
To evaluate if and how the simulated echellograms can be handled by a pipeline, the ELT-HIRES DRS group applied the beta recipes of the \crires\ DRS package. 
In spite of ELT-HIRES being fiber-fed and \crires\ a true slit spectrograph
\footnote{Details on \crires\ and its DRS can be found at \url{https://www.eso.org/sci/facilities/develop/instruments/crires\_up.html}}
, they share enough similarities in the echellograms (high-resolution, cross dispersed, curved spectral orders with tilted pseudo-slit projection), so that the recipes could be applied and tested without relevant modifications.
The first step was to demonstrate the capability of the E2E simulator to preliminary close the loop with the DRS, since the delivered frames of the simulator need to act as input for this pipeline.
The strong link between the E2E and DRS is crucial especially in future E2E simulator versions, since this will allow us to characterize the performance of the instrument by translating directly a opto-mechanical architecture into scientific terms, by simulating all the main ELT-HIRES science cases. 

\begin{figure*}
\centering
{\includegraphics[width=0.95\textwidth, height=1\textheight, keepaspectratio] {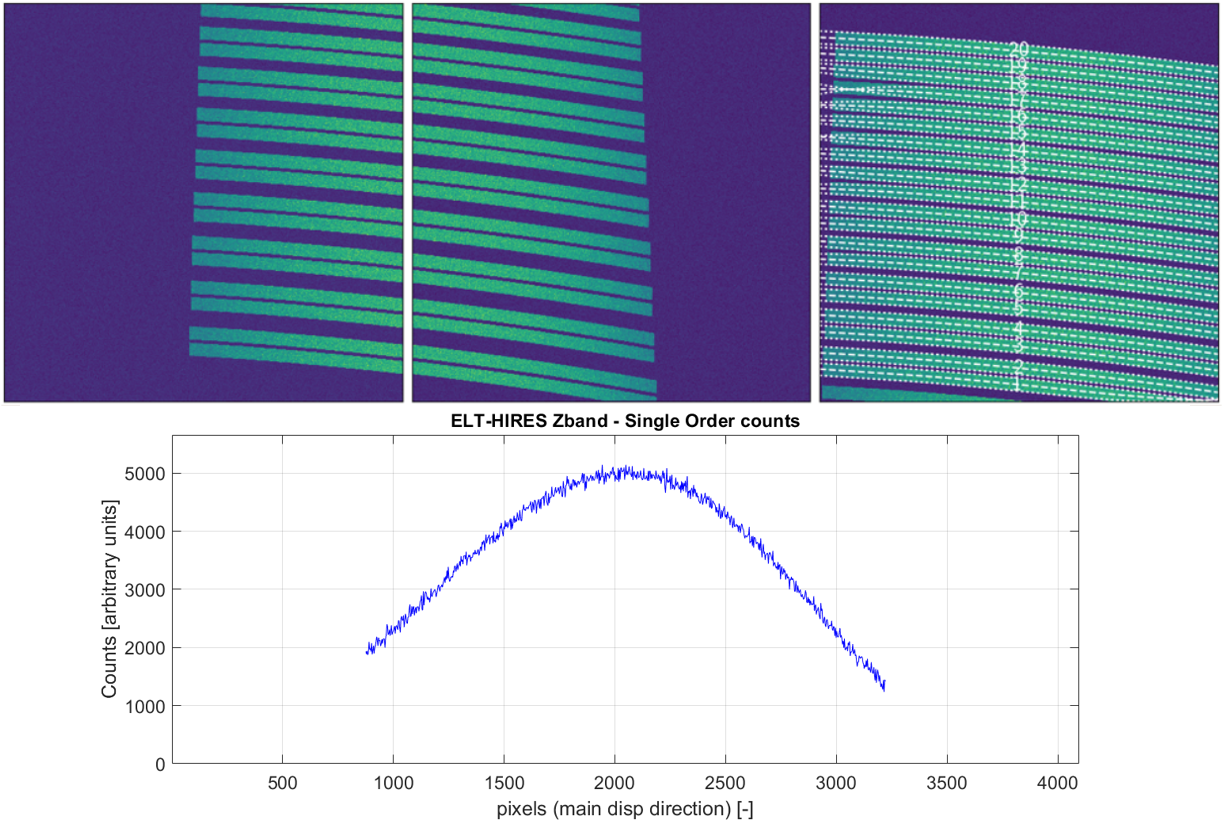}}
\caption{\footnotesize
ELT-HIRES Z Band flat field frame produced with the E2E simulator. \emph{Left}: RAW data. \emph{Right}: order tracing from \crires\ pipeline. The derived polynomials and order number are plotted in white. The dashed lines represent the mid-line of the orders, the dotted ones the top and bottom, derived via smoothing, thresholding and edge-detection.
\emph{Bottom}: Optimally extracted 1D-spectrum of a single arbitrarily chosen spectral order. The spectrum shows the characteristic shape of the blaze function, which is computed considering the diffraction, interference and shadowing effect from the single grating grooves.}
\label{fig:E2E_24_DRS_1}
\end{figure*}

\begin{figure*}
\centering
{\includegraphics[width=1\textwidth, height=1\textheight, keepaspectratio] {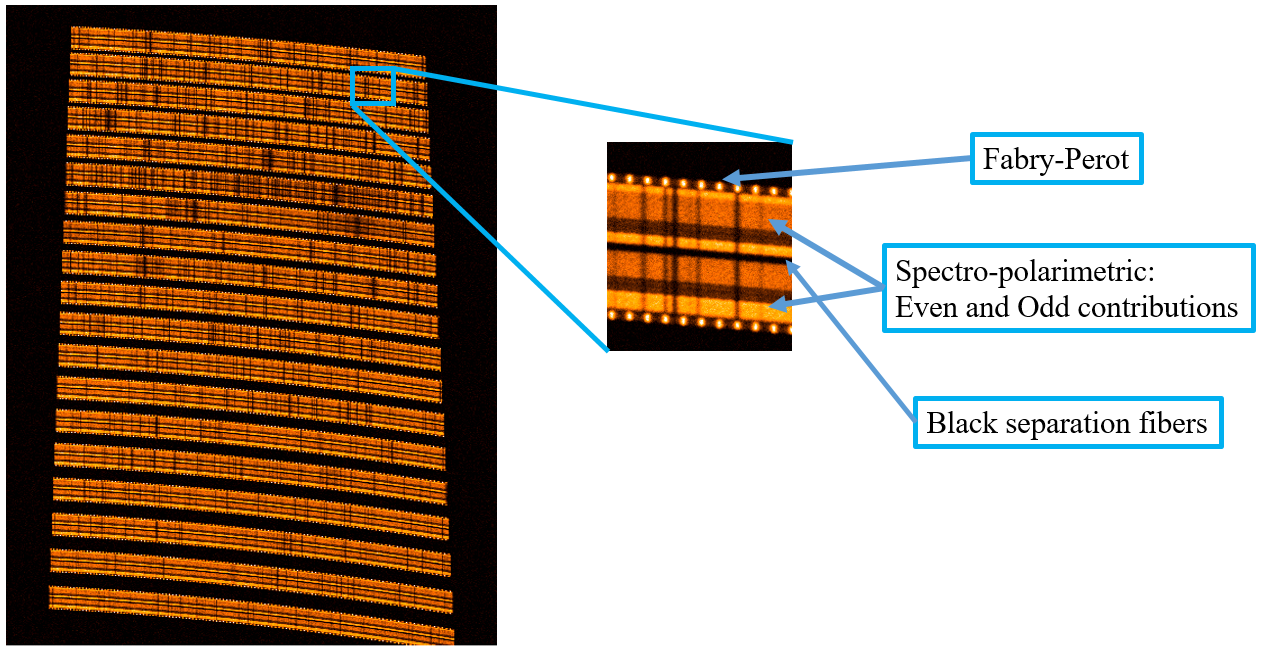}}
\caption{\footnotesize
ELT-HIRES K8V star (with $m_{V} = 10$) simulated Z-band raw frame. Spectro-polarimetric mode, with Even and Odd contributions in the two fibers-bundles creating the entrance fiber-slit and separated by a set of black fibers. In the zoomed-in window the different illumination of the vertically re-arranged fibers is clearly visible. Simultaneous Fabry-Perot reference spectra are on the top and bottom of the spectral order.}
\label{fig:E2E_24_DRS_2}
\end{figure*}

Two different raw frames (plus trivial ones like BIAS, not reported here) were produced with the E2E simulator.
First, a flat-field frame in the Z-band (830 - 950 nm), top-left panel of Fig.\ref{fig:E2E_24_DRS_1}. 
Second, a scientific Z-band exposure of a K8V star with $m_{V} = 10$ and exposure time $50\mathrm{s}$ observed with a seeing of $0.85^{\prime\prime}$, which produces differently illuminated fiber-slits, according to the specific position of each fiber in the bundle (fibers in the center receive more flux than fibers at the edge).
We chose polarimetric mode, where both target and sky bundle are illuminated by the same object, with simultaneous Fabry-Perot reference spectra on the top and bottom of the spectral order. 
The instrumental efficiency of the spectrograph was computed by estimating the transmission of glasses (assuming flat AR-coatings contribution) and through analytical calculation of the echelle grating blaze profile for each order, considering theoretical terms for diffraction, interference and shadowing for the single grooves (see e.g. \cite{Ref-Eversberg}).
The resulting echellogram is shown in Fig.\ref{fig:E2E_24_DRS_2}.

These frames are passed to the DRS of \crires, in active development phase at the time.
The tracing of spectral orders, i.e. finding their locations on the detector and fitting them with polynomials, was carried out successfully, as can be seen in the top-right panel of Fig.\ref{fig:E2E_24_DRS_1}. The recovered flat-field-lamp spectrum in 1D shows the characteristic blaze function shape, indicating a correct flux extraction. Perfectly vertical alignment of the fiber-slit to pixel columns was assumed, since this is how the simulations were made.

\begin{figure*}
\centering
{\includegraphics[width=0.95\textwidth, height=1\textheight, keepaspectratio] {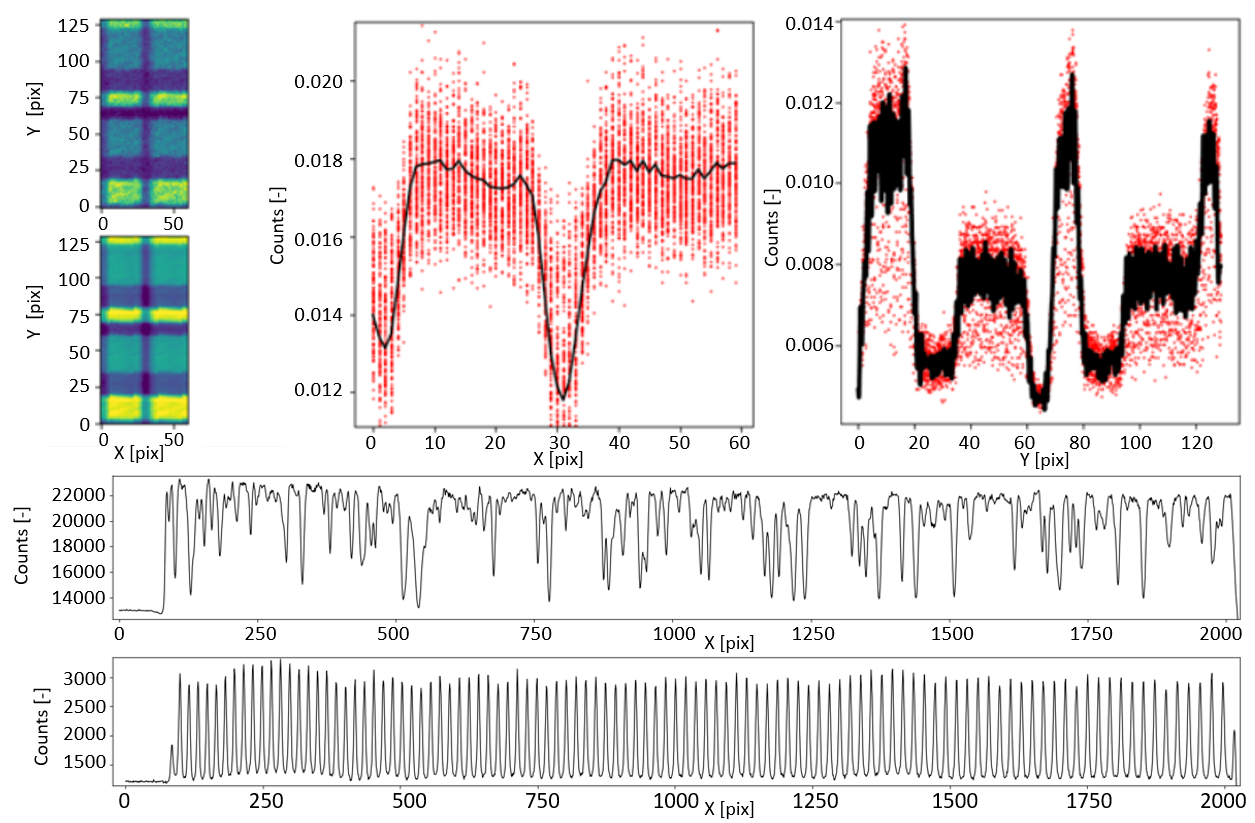}}
\caption{\footnotesize
Optimal extraction in order 146 of HIRES Z-band. \emph{Top-left}: Cut-out of the raw data from E2E-simulations, and its model (below) as reconstructed from the 1D spectrum and 1D slit-illumination.
\emph{Top-middle}: Snap-shop of the extraction step in main dispersion direction. In red we show the individual pixel values projected onto the
spectral dimension, and in black the resulting 1D-spectrum. Note that the two strong absorption lines match those in the raw frame cut-out.
\emph{Top-right}: The slit-illumination profile for the same region, again
with individual pixel values in red and the resulting 1D profile in black.
The profile matches the features that the raw frame shows in vertical direction.
\emph{Bottom panels}: Extraction results for the scientific spectrum and the etalon reference spectrum.}
\label{fig:E2E_24_DRS_3_4}
\end{figure*}

The spectra from science frame were similarly extracted by slit decomposition (\cite{Ref-Piskunov}), a robust algorithm that is independent of the slit illumination function, meaning that differently illuminated fibers along the pseudo-slit can be extracted together.
As can be noticed from top-middle and top-right panels of Fig.\ref{fig:E2E_24_DRS_3_4}), the spectrum profile along the main dispersion direction and the illumination profile of the fiber-slit in cross-dispersion direction were properly retrieved. The reconstructed 2D-model matches the cut-out of the raw frame in all relevant aspects (top-left panels).

We chose a single order of the science frame for analysis and applied the order-trace polynomials from the flat-field frame with different vertical offsets and heights (in pixels) to select the pseudo-slit and the etalon spectra separately for extraction. Extracting these two 1D-spectra keeps them on identical pixel scales so that the wavelength-scale can be transferred from one to the other, once the etalon lines are measured and calibrated. Note that this last step is not carried out here.

For exercise, we tried to calibrate the one-dimensional spectrum by assuming a constant dispersion across the order. Although this is a very rough and preliminary approximation, we recovered the initial radial velocity target by use of cross correlation technique (see e.g. \cite{Ref-Pepe-CCF}). 

Finally, we comment on the computational efficiency and cost obtained for these simulation. As in the case of ESPRESSO, we fired up a cluster of 10 \texttt{g2x.large} EC2 instances that completed the simulation of the $Z$ band of E-ELT HIRES in $\sim$ 10 hrs.

\section{Summary and Future perspectives}
\label{sec:Conclusions}
We presented in this paper a modular and scalable End-to-End simulator targeted to the  modeling of fiber-fed high resolution spectrographs and characterized by a flexibility level that allows it to be used to emulate a wider range of spectrographs. 
The adoption of high parallelism at the level of the simulations, and the introduction of distributed computation via Cloud Computing, allowed us to simulate data for two high resolution spectrographs ESO-ESPRESSO and ELT HIRES. The presented approach permitted to reduce by orders of magnitude the requested computational time to render the raw frames while keeping a fine-grained level of accuracy. 
The simulated data, in the case of ESO-ESPRESSO, has been compared to actual frames from the instrument. We were able to reproduce the overall spectral format and the LSF of the spectrograph at the level of few ppm while, for ELT-HIRES, we demonstrated the applicability of our approach by exploiting the simulator to test the re-usability Data Reduction Software as a baseline for current phase of the instrument.

Concerning the future perspective of our tool, we are currently integrating the simulator outputs with ESPRESSO-DRS pipeline with the final purpose of executing complete simulations and to directly translate the effect of different instrumental parameters (e.g. thermal detector instabilities) into scientific performance. 
In addition to this, we plan to extend the simulation chain with models of other effects, such as thermo-mechanical instabilities (of the telescope, instrument and detector), telescope-pointing residual and secondary guiding errors. 
All these improvements will be fundamental for deeply characterising different operational scenarios and a series of systematics experienced in ESPRESSO that will be illustrated in a forthcoming paper.

\begin{acknowledgements}
We thank the anonymous referee for his/her very constructive and helpful report.
\\M. G. and M. L. gratefully thank V. S.,  M. C. and G. M. for huge support during the preparation of this manuscript.
\end{acknowledgements}


%



\end{document}